\documentclass[12pt]{article}
\usepackage{geometry}                % See geometry.pdf to learn the layout
\usepackage[latin1]{inputenc}
\usepackage[T1]{fontenc}
\usepackage{graphicx}
\usepackage{amssymb}
\usepackage{epstopdf}

\title{Kowalevski's analysis of the swinging Atwood's machine.}
\author{O. Babelon\footnote{L.P.T.H.E. Universit\'e Pierre et Marie
Curie--Paris6; CNRS; UMR 7589; Bo\^{\i}te 126, Tour 24,  $5^{eme}$ \'etage,
 4 place Jussieu, F-75252 PARIS CEDEX 05}, M. Talon$^{*}$ and M. Capdequi
Peyranère\footnote{L.P.T.A., Université Montpellier II; CNRS/IN2P3;  UMR 5207;
Case Courrier 070, Bât. 13, place Eugène Bataillon, 34095 Montpellier Cedex 5}}
\date{September 2009}

\begin{document}
\maketitle
\abstract{We study the Kowalevski expansions near singularities of the swinging
Atwood's machine. We show that there is a infinite number of
mass ratios $M/m$ where such expansions exist with the maximal number
of arbitrary constants. These expansions are of the so--called weak Painlevé
type. However, in view of these expansions, it is not possible to distinguish
between integrable and non integrable cases.}

\newpage
\section{Introduction}

\begin{figure}[hbtp]
\begin{center}
\includegraphics[height= 7cm]{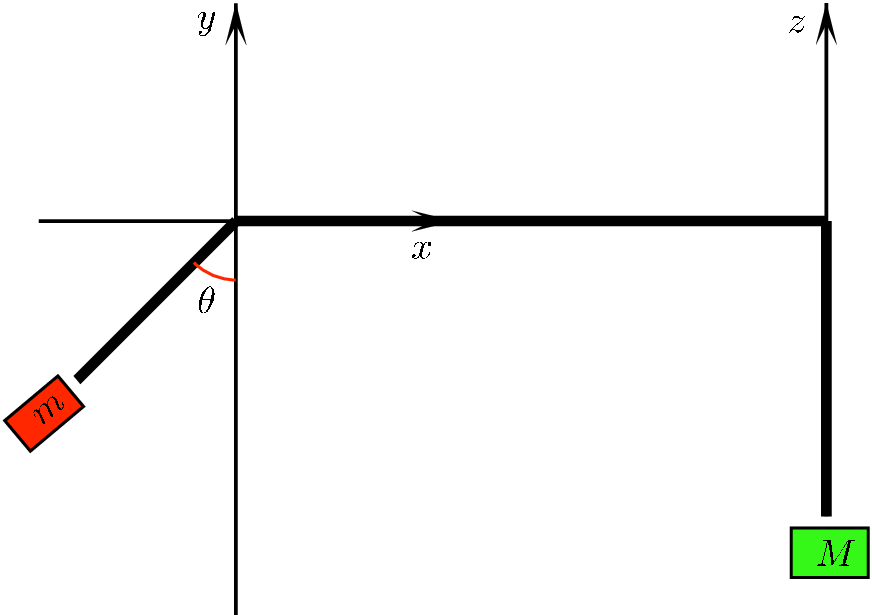}
\caption{Swinging Atwood's machine.}
\label{figatwood}
\end{center}
\nonumber
\end{figure}

The swinging Atwood's machine is a variable length pendulum of mass
$m$ on the left, and a non swinging mass $M$ on the right, tied together by a string, in a
constant gravitational field, see Figure~(\ref{figatwood}). The
coupling of the two masses is expressed by the fact that the length of the
string is fixed:
$$
\sqrt{x^2 + y^2} + |z| = L, \quad \Longrightarrow x^2 + y^2 = (|z|-L)^2
$$
Up to a choice of origin for z, one can assume $L=0$, so the constraint is the
cone $z^2=x^2+y^2$. To describe the dynamics we choose to work with constrained
variables and write a Lagrangian 
$$
{\cal L} = {m\over 2}(\dot{x}^2 + \dot{y}^2 ) + {M\over 2} \dot{z}^2 - g(m y+ M
z) +{ \lambda\over 2} (x^2 + y^2 - z^2)
$$
where $\lambda$, a Lagrange multiplier
(of dimension $MT^{-2}$), has been introduced, whose equation of
motion enforces the constraint.
The equations of motion read :
\begin{eqnarray}
m \ddot{x} &=& \lambda x \label{eqx}\\
m \ddot{y} &=&- m g+  \lambda y \label{eqy}\\
M \ddot{z} &=&- M g-  \lambda z \label{eqz}\\
0 &=& x^2 + y^2 - z^2
\end{eqnarray}

From these equations one can express $\lambda$ in terms of positions and
velocities:
\begin{equation}
\lambda = { x \ddot{x} + y \ddot{y} - z\ddot{z} +g (y-z) \over {1\over m}
(x^2+y^2)+{1\over M} z^2} 
={mM\over M+m}{ \dot{z}^2 - \dot{x}^2-\dot{y}^2 +g (y-z) \over  z^2}
\label{eqlambda}
\end{equation}

Alternatively, rescaling 
$$
x\to {1\over \sqrt{m}} x,\quad y\to {1\over \sqrt{m}} y,\quad z\to {1\over \sqrt{M}} z
$$
we can view the system as a unit mass particle moving on a  cone
$$
z^2 = {M\over m} (x^2+y^2)
$$
subjected to a constant  field force
$$
\pmatrix{f_x\cr f_y\cr fz} = \pmatrix{0\cr -g\sqrt{m} \cr -g \sqrt{M}}
$$
The slope of the force in the $(y,z)$ plane coincides with the angle of the
cone.

The swinging Atwood's machine has been studied in great detail by N. Tufillaro and
his coworkers, see [1--7]. They have first studied numerically the equations of
motion and shown that for most values of the mass ratio $M/m$ the motion
appears to be chaotic, however for some values, like 3, 15, etc. the motion
seems less chaotic and could perhaps be integrable. In a further study, Tufillaro~\cite{T86}
showed that the system is indeed integrable for $M/m=3$ by exhibiting a change
of coordinates, somewhat related to parabolic coordinates, in which 
separation of variables occurs. He was then able to solve the equations of motion
in terms of elliptic functions, which is quite peculiar since in general
integrable systems with two degrees of freedom can be solved only in terms of
hyperelliptic functions, such as for the Kowalevski top~\cite{Kowa}. He also
obtained the second conserved quantity which ensures integrability. In the same
paper, he conjectured that the system is integrable for
$M/m=15,\cdots, 4n^2-1$, with $n$ integer.

However, later on, Casasayas, Nunes and Tufillaro proved~\cite{CNT90} that the system
can be integrable for discrete values of the ratio $M/m$ only in the interval
$]1,3]$, using non integrability theorems developed by
Yoshida~\cite{Yos} and Ziglin. The essence of the Yoshida--Ziglin argument is
to study the monodromy developed by Jacobi variations around an exact
solution, when the time variable describes a loop in the complex plane.
The monodromy  must preserve conserved quantities, but  this is
impossible in general if the monodromy group is not abelian. In the case at hand
one can compute monodromies from hypergeometric equations and conclude. We have also 
been informed by private communication of J.P. Ramis, that he and his coworkers have proven 
that the swinging Atwood's machine is never integrable except for $M/m=3$, using 
methods from differential Galois theory.

The aim of our paper is to work out the Kowalevski analysis for this model.
Let us recall the idea of the Kowalevski method. If a dynamical system is
{\em algebraically} integrable one can expect to obtain expressions for the
dynamical variables in terms of quotients of theta functions defined on the
Jacobian of some algebraic curve of genus g, where g=2 for a system with 2
degrees of freedom. Only quotients may appear because theta functions have
monodromy on the Jacobian torus, which needs to cancel. Hence denominators which
can vanish for any given initial conditions and for some finite value, in
general complex, of time will appear in the solution.
Hence the equations of motion must admit Laurent solutions --that is divergent
for some value of time, with as many parameters as there are initial conditions.
S. Kowalevski first noted~\cite{Kowa}, that this imposes strong constraints on
these equations, from which she was able to deduce the celebrated Kowalevski
case of the top equation.

Looking for Laurent solutions to the swinging Atwood's machine equations of
motion in the integrable case $M/m=3$ we first noted that there are none, but there
exists so-called weak Painlevé solutions, that is Laurent developments not in
the time variable $t$ but in some radical $t^{1/k}$, generally called Puiseux
expansions.

It had already been discovered by A. Ramani and coworkers~\cite{Ram} that some
integrable systems require weakening the Kowalevski--Painlevé analysis to obtain
expansions at infinity of dynamical variables. This may be explained in general,
and is certainly the case for our example, by the fact that there is a
``better'' variable which has true Laurent expansions and time itself can be
expressed in terms of this variable through an algebraic equation which happens
to produce the given radicals. Moreover Ramani et al. advocated the idea that
the existence of weak Painlevé solutions is a criterion of integrability, like
in the Kowalevski's case. 

For our model of the swinging Atwood's machine, we find
that there are weak Painlevé  solutions not only when $M/m=15$ 
but for a whole host of other values of the mass
ratio, all of them corresponding to obviously non integrable cases. Hence this
model provides a large number of counterexamples to the above idea. We then
study in detail the solutions around infinity which can be extracted from
these Kowalevski developments. Using Padé approximants we are able to extend
these solutions beyond the first new singularity and observe how the new
singularities obey Kowalevski exponents.

We also comment on the Poisson structure of the model, which is interesting
due to the constraints between the dynamical variables, and the Poisson
brackets of the variables appearing in the Laurent series, which happens to be
of a nice canonical form. We notice that this illustrates the fact that it is
the global character of the conserved quantities that is of importance in
defining an integrable system.

One of us (M.T.) is happy to acknowledge useful conversations with J.P. Ramis
and J. Sauloy from Toulouse University, about their work on differential Galois
theory applied to the swinging Atwood's machine. Finally we are happy to thank
the Maxima team\footnote{http://maxima.sourceforge.net/} for their software, with
which we have performed the computations in this paper. 

\section{Hamiltonian setup.}

The description we have given of the swinging Atwood's machine is  a
constrained system in the Lagrange formulation,
so that the equations of motion take a nice algebraic form.

In the articles [1--7] polar coordinates are used, so the constraint is
``solved'' but the price to pay is the use of trigonometric functions.
Using polar coordinates $x=r\sin \theta$,
$y=-r\cos\theta$ the Hamiltonian
reads:
\begin{equation}
H={1\over 2(m+M)}p_r^2 +{1\over 2m r^2}p_{\theta}^2 
+gr(M-m\cos\theta)
\label{hampol}
\end{equation}
where $p_r=(m+M)\dot{r}$ and $p_ \theta=m r^2\dot{\theta}$.

We now give a Hamiltonian description of this system, using as
dynamical variables the three coordinates $x, y, z$ and
the three momenta $p_x, p_y, p_z$ with canonical Poisson brackets.
The constraint
\begin{equation}
C_1 \equiv z^2-x^2-y^2 =0
\label{const}
\end{equation}
generates the flow:
\begin{equation}
\{C_1,p_x\} = -2x,\quad \{C_1,p_y\} = -2y,\quad \{C_1,p_z\} = 2z
\label{moment}
\end{equation}
which is also  generated by the one parameter 
group acting on phase space by:
$(x,y,z)\to(x,y,z),\quad (p_x,p_y,p_z)\to(p_x-\mu x,p_y-\mu y,p_z+\mu z)$
where $\mu$ is the group parameter.

We want to describe the dynamics of our model as a Hamiltonian system 
obtained by reduction of an invariant system under this group action~\cite{BBT}.
In order to do that, consider the functions:
\begin{eqnarray*}
A_x &=& z p_y + y p_z \\
A_y &=& z p_x + x p_z \\
A_z &=& x p_y - y p_x
\end{eqnarray*}
These functions Poisson commute with the constraint $C_1$ hence are invariant
under the group action. They are not independent however, since they are related
by:
\begin{equation}
 y A_y -x A_x + z A_z = 0
\label{C2}
\end{equation}

It is easy to check the Poisson brackets:
$$
\{ A_x,A_y\} = -A_z,\quad \{A_x,A_z\} = -A_y,\quad  \{A_y,A_z\} = A_x
$$
$$
 \{ A_x,x\} =0,\quad \{ A_x,y\} =z, \quad \{ A_x,z\} =y 
$$
$$
 \{ A_y,x\} =z,\quad \{ A_y,y\} =0, \quad \{ A_y,z\} =x 
$$
$$
 \{ A_z,x\} =-y,\quad \{ A_z,y\} =x, \quad \{ A_z,z\} =0 
$$

Let us consider the invariant Hamiltonian:
\begin{equation}
H={1\over 2(m+M)z^2}\left[ A_x^2+A_y^2 + {M\over m}A_z^2 \right] + Mgz +mgy
\label{hamred}
\end{equation}

To check that $H$ generates the equations of motion on the reduced
system, we compute:
\begin{eqnarray}
\dot x = \{H,x\} &=& {1\over m+M} {1\over z^2} \left(zA_y-{M\over
m}yA_z\right)\label{xdot}\\
\dot y = \{H,y\} &=& {1\over m+M} {1\over z^2} \left(zA_x+{M\over
m}xA_z\right)\label{ydot}\\
\dot z = \{H,z\} &=& {1\over m+M} {1\over z^2} \left(xA_y +yA_x\right)
\label{zdot}
\end{eqnarray}
The right hand sides of these equations are linear in the momenta $p_x$, $p_y$,
$p_z$, however we cannot invert the system uniquely in order to express the
momenta in terms of the velocities. This is
because, due to the symmetry ($\{H,C_1\}=0$) we have
$x\dot x+y\dot y=z\dot z$ so the equations are not independent. The solution is:
\begin{eqnarray}
p_x &=& m\dot x +\mu x\label{px}\\
p_y &=& m\dot y +\mu y\label{py}\\
p_z &=& M\dot z -\mu z \label{pz}
\end{eqnarray}
where $\mu$ is arbitrary.
Similarly we compute $\ddot x=\{H,\dot x\}$, etc... where $\dot x$, etc... are
the right hand sides of the above equations. Performing this calculation and
using the constraint $C_1$ and eq.(\ref{C2}), we obtain
the Lagrangian equation of motion (\ref{eqx}--\ref{eqz}),  with $\lambda$
given by:
$$\lambda={mM\over m+M}{1\over z^2}\left[
g(y-z)-{1\over m^2z^2}A_z^2\right]$$
This coincides with eq.(\ref{eqlambda}), as can be cheked using again
eqs.(\ref{px}--\ref{pz}) and the constraint $C_1$, to express $A_z$ in terms of
$\dot x$, $\dot y$, $\dot z$.

Finally we express the energy in terms of velocities still using the
constraints. We find:
\begin{equation}
E = {m\over 2}(\dot{x}^2 + \dot{y}^2 ) + {M\over 2} \dot{z}^2 + g(m y+ M z)
\label{energy}
\end{equation}
which agrees with what we expect from the Lagrangian formulation.

\section{The integrable case.}

In order to understand what sort of Laurent expansions appears in the model it
is useful to first consider the case $M/m=3$ which has been integrated by
Tufillaro~\cite{T86}. Let us recall some of his results. He discovered that 
using polar coordinates $(r,\theta)$ such that $x=r\sin\theta$,
$y=-r\cos\theta$ and $r=z$, and setting:
$$ \xi^2 = z[1 + \sin (\theta/2)],\quad
\eta^2=z[1 -\sin(\theta/2)]$$
then the Hamilton--Jacobi equation separates in the variables $(\xi,\eta)$. These
look like parabolic coordinates, except that the half--angle $\theta/2$ is used.
Knowing $\xi$ and $\eta$ one can recover $x$ and $y$ by:
\begin{equation}
x_\pm\equiv x\pm i y = \pm {i\over 2} {(\xi \mp i \eta)^3 \over (\xi\pm i
\eta)},\quad z={1\over 2}(\xi^2+\eta^2)
\label{xxietaz}
\end{equation}
In fact, just for $M=3m$, two terms involving couplings
between $\xi$ and $\eta$ 
disappear, and one gets, with momenta $p_\xi=4\dot\xi(\xi^2+\eta^2)$ etc. the
expression of the Hamiltonian, in which we have set $m=1$:
$$H = [(p_\xi^2+p_\eta^2)/8 + 2g(\xi^4+\eta^4)]/(\xi^2+\eta^2)$$

Then it is clear that in this case the action $S$ separates as a sum $S_\xi(\xi)+
S_\eta(\eta)$ where $S_\xi$ and $S_\eta$ obey different elliptic equations
(corresponding to different elliptic moduli):
\begin{eqnarray}
(\partial_\xi S_\xi)^2 &=& -16 g \xi^4 + 8E\xi^2  +I \equiv P_+(\xi)
\label{Sxi}\\
(\partial_\eta S_\eta)^2 &=& -16 g \eta^4 + 8E\eta^2 -I \equiv P_-(\eta)
\label{Seta}
\end{eqnarray}
where I is the separation constant. It can be expressed in terms of dynamical
variables by substracting the above two equations multiplied resp. by 
$\eta^2$ and $\xi^2$, which eliminates $E$. Moreover we replace:
\begin{equation}
\partial_\xi S =p_\xi=4\dot\xi(\xi^2+\eta^2),\quad 
\partial_\eta S =p_\eta=4\dot\eta(\xi^2+\eta^2)
\label{momS}
\end{equation}
We get:
$$I/16= (\xi^2+\eta^2)(\eta^2\dot\xi^2-\xi^2\dot\eta^2)
+g\xi^2\eta^2(\xi^2-\eta^2)/(\xi^2+\eta^2)$$
Returning to polar coordinates the integral of motion takes the form:
$$I/16=r^2\dot\theta\big[ \dot r \cos(\theta/2)-{r\dot\theta\over
2}\sin(\theta/2)\big] + gr^2\sin(\theta/2)\cos^2(\theta/2)$$

We want to see if the equations of motion admit a solution which diverges
at finite time,  and in that case what is the behavior of the Laurent expansion.

The general solution of the Hamilton--Jacobi equation is:
$$S= -Et+\int^\xi \sqrt{P_+(\xi)}\,d\xi +\int^\eta \sqrt{P_-(\eta)}\,d\eta$$
According to the general theory we get the solution of the equations of motion
by writing $\partial_E S = c_E$ and $\partial_I S = c_I$ for two constants
$c_E$ and $c_I$. For $c_I\neq 0$ we get:
\begin{eqnarray}
 t+c_E &=& \int^\xi {4\xi^2\over\sqrt{P_+(\xi)}}d\xi + \int^\eta
{4\eta^2\over\sqrt{P_-(\eta)}}d\eta\label{ellipt1}\\
c_I &=& {1\over 2}\int^\xi {1\over\sqrt{P_+(\xi)}}d\xi - {1\over 2}\int^\eta
{1\over\sqrt{P_-(\eta)}}d\eta\label{ellipt2}
\end{eqnarray}

For $I=0$, the elliptic integrals degenerate to trigonometric ones. We get:
\begin{equation}
 t+c_E = -{1\over 2
\omega}\left(\sqrt{1-\alpha\xi^2}+\sqrt{1-\alpha\eta^2}\right),~\alpha=2g/E,
~\omega=g/\sqrt{2E}
\label{xietat}
\end{equation}
\begin{equation}
 {1-\sqrt{1-\alpha\xi^2}\over
1+\sqrt{1-\alpha\xi^2}}=K^2{1-\sqrt{1-\alpha\eta^2}\over
1+\sqrt{1-\alpha\eta^2}},\quad K^2=e^{c_I}
\end{equation}
so that setting
$\xi=\sin(\phi_\xi)/\sqrt{\alpha},\;\eta=\sin(\phi_\eta)/\sqrt{\alpha}$ the
second equality reads:
$$\tan(\phi_\xi/2)=K \tan(\phi_\eta/2)$$
Using the variable $s=\tan(\phi_\xi/2)$,
$\xi$ and $\eta$ can be expressed rationally:
$$\xi={1\over\sqrt{\alpha}}{2s\over 1+s^2},\quad
\eta={1\over\sqrt{\alpha}}{2Ks\over K^2+s^2}$$

Finally, one gets the time variation of $S\equiv s^2$ by using
eq.(\ref{xietat}) which implies:
$$ \omega dt =  dS \left[{1\over (1+S)^2}
+{K^2\over(K^2+S)^2}\right]= -{8iK\over K^2-1}{U\,dU\over (U^2-1)^2}$$
where we have parametrized $S$ as:
$$
S= iK\,{(K+i)U+(K-i)\over (K-i)U-(K+i)}
$$
The variable $U$ has been defined to send the poles $S=-K^2$ and $S=-1$ to
$U=\pm 1$. One gets the two parameters solution (parameters $K$ and $E$)
up to an origin for time, which we fix by requiring that $t=0$ for $U=0$:
$$
U^2= {t\over t-t_\infty}~{\rm or}~U^2-1={t_\infty\over t -t_\infty}
\Longrightarrow t=-t_\infty{ U^2\over
1-U^2},\quad t_\infty={1\over\omega}{4iK\over K^2-1}
$$
We shall soon see that $t=t_\infty$ is a second singularity of the dynamical
variables, that we can express explicitly. For ease of comparison
with the following, we present $x_\pm(t)=x(t)\pm i y(t)$:
\begin{eqnarray*}
 x_+&=& -{2Kg\over \omega^2}\;{\left[(K-i)U -K-i\right]\,\left[(K+i)U+K-i\right]
 \over{(K^2-1)^2(U^2-1)^2}}\,{1\over U}\\
x_-&=& {2Kg\over \omega^2}\;{\left[(K-i)U-K-i\right]\,\left[(K+i)U+K-i
 \right]\over{(K^2-1)^2(U^2-1)^2}}\, U^3\\
 z&=& i{2Kg\over \omega^2}\;{\left[(K-i)U-K-i\right]\,\left[(K+i)U+K-i
 \right]\over{(K^2-1)^2(U^2-1)^2}}\, U \\
 \lambda&=&-{3\omega^2\over 64 K^2}{(K^2-1)^2(K^2+1)(U^2-1)^5\over
 \left[(K-i)U -K-i\right]\,\left[(K+i)U+K-i\right]U^4}
\end{eqnarray*}

In terms of the $t$ variable, we get the simpler expressions:
\begin{eqnarray}
 x_+(t) &=& -{2Kg\over \omega^2(K^2-1)^2} \left[ (K^2+1)
{\left({t-t_\infty\over t_\infty}\right)^{3/2}}\left({t_\infty\over
t}\right)^{1/2}-4iK\left({t-t_\infty\over t_\infty}\right)^2\right] 
\nonumber\\
&&\label{xplust}\\
x_-(t) &=& {2Kg\over \omega^2(K^2-1)^2} \left[ (K^2+1)\left({t\over
t_\infty}\right)^{3/2}\left({t_\infty\over t- t_\infty}\right)^{1/2}
-4iK\left({t\over t_\infty}\right)^2\right]
\nonumber\\
&&\label{xmoinst}
\end{eqnarray}

We see that $x_+$ behaves as $t^{-{1\over 2}}$ and $x_-$ behaves as
$t^{3\over 2}$ when $t\to 0$.
If we expand around $t=0$ we get Puiseux expansions in $t^{1\over 2}$.
These expansions depend on three parameters, $K$ and $E$ plus the
origin of time $t_0$. This is because we are analyzing the trigonometric
solution which fixes one of the constants to $I=0$. We shall see later on that
it can be generalized to a four parameter expansion in the elliptic case.
The energy parameter appears factorized in front of $x_+$ and $x_-$ in the form
of $g/\omega^2=2E/g$. 

Around $t_\infty$, we see that $x_+$ behaves as $(t-t_\infty)^{3\over 2}$
and $x_-$ behaves as $(t-t_\infty)^{-{1\over 2}}$ which is symmetrical with the
behaviour at $t=0$. This is compatible with the fact that the equations of
motion admit a symmetry 
$(x_+(t),x_-(t)) \leftrightarrow (-x_-(t),-x_+(t))$.

Remark that $x_\pm(t)$ are defined on the 
two sheeted covering of  the Riemann
sphere with two branch points at $t=0$ and $t=t_\infty$.
The variable $U$ that we have introduced is in fact a uniformizing variable for
this covering, so that $x_\pm(t)$ are rational functions of $U$. Moreover 
$U \leftrightarrow -1/U$ corresponds to $t \leftrightarrow (t_\infty -t)$ and
exchanges $x_+$ and $-x_-$. The extra minus sign means that we have to change
the determination of the square root in the $t$ variable. The $U$ variable
makes this completely unambiguous:
$$ x_+\left(-{1\over U}\right)=-x_-(U),\quad z\left(-{1\over
U}\right)=z(U),\quad \lambda\left( -{1\over U}\right)=\lambda(U)$$
We emphasize that, although the system is integrable, the solutions diverge
with {\em square root} singularities at finite times $t=0$, and $t=t_\infty$.

We now return to the elliptic case. Let us define the variables
$X=\xi^2-E/(6g)$  and $Y=\eta^2-E/(6g)$. The equations
(\ref{ellipt1},\ref{ellipt2}) become:
\begin{eqnarray*}
 t+c_E &=& {1\over 4i\sqrt{g}} \int^X {(X+E/(6g))
dX\over\sqrt{P_+(X)}} + {1\over 4i\sqrt{g}} \int^Y {(Y+E/(6g))
dY\over\sqrt{P_-(Y) }} \\
c_I &=& {1\over 4i\sqrt{g}} \int^X {
dX\over\sqrt{P_+(X)} } - {1\over 4i\sqrt{g}} \int^Y {
dY\over\sqrt{P_-(Y)} }
\end{eqnarray*}
where now:
$$P_\pm(X) = 4X^3-g_2(\pm I)X-g_3(\pm I)$$ $$
g_2(I)={1\over 3g^2}\left(E^2+{3\over 4}gI\right),\quad g_3(I)=
{E\over 27 g^3}\left(E^2+{9\over 8}gI\right)$$
Introducing the Weierstrass  functions
$$X=\wp_1(Z_1)\equiv \wp(Z_1,g_2(I),g_3(I)), \quad
Y=\wp_2(Z_2)\equiv \wp(Z_2,g_2(-I),g_3(-I))$$
the above integrals reduce to:
\begin{eqnarray}
 t+c_E &=& {1\over 4i\sqrt{g}} \left[ {E\over 6g} (Z_1+Z_2) -\zeta_1(Z_1)
-\zeta_2(Z_2)\right]\label{ellipt3}\\
c_I &=& {1\over 4i\sqrt{g}} \left[ Z_1-Z_2 \right]\label{ellipt4}
\end{eqnarray}
where $\zeta$ is the Weierstrass  zeta function, $\zeta'=-\wp$. The $\wp$
function has two periods 2$\omega_j$, $j=1,2$, so that $\wp(z+2\omega_j)=
\wp(z)$, but the zeta function is quasi periodic, $\zeta(z+2\omega_j)=
\zeta(z)+2\eta_j$. Here we have two set of periods $\omega_j$ and $\eta_j$ 
according to the function $\wp_1$ or $\wp_2$, which are in fact functions
$\omega_j(\pm I)$ and $\eta_j(\pm I)$.

Note that $x_\pm(t)$ have poles and zeroes when $\xi\pm i \eta$ vanish, that
is when $\xi^2+\eta^2=X+Y+E/(3g)=0$. Hence we have to solve:
\begin{eqnarray}
&&E/(3g)+ \wp_1(Z_1)+\wp_2(Z_2) = 0 \label{ellpole1}\\
&&Z_1-Z_2 -4i\sqrt{g} c_I = 0 \label{ellpole2}
\end{eqnarray}
But differentiating
eqs.(\ref{ellipt3},\ref{ellipt4}) we find $\delta Z_2=\delta Z_1$ and
$$\delta t ={1\over 4i\sqrt{g}}\left({E\over 3g}+ \wp_1(Z_1)+\wp_2(Z_2)\right)
\delta Z_1 +{1\over 8i\sqrt{g}}\left(\wp_1'(Z_1)+\wp_2'(Z_2)\right)(\delta
Z_1)^2+\cdots$$
The first term vanishes when $\xi^2+\eta^2=0$ hence around such a zero
$\delta Z_1\simeq\sqrt{\delta t}$. As a consequence, in view of
eq.(\ref{xxietaz}),
$x_\pm(t)$ behaves as either $\delta t^{-1/2}$ or $\delta t^{3/2}$ at such a
point, according to the vanishing of $\xi+i\eta$ or $\xi-i\eta$.
Note this is similar to the trigonometric case.

However finding the pattern of these singularities is messy, because in the
equations~(\ref{ellpole1},\ref{ellpole2})
we have two incommensurate lattices of periods for the two Weierstrass
functions. However we can easily see that there is an infinite number of
singularities. This is because since the two lattices are incommensurate, for
any large $R$ and small $\epsilon$, one can choose $V$ in the first lattice and
$W$ in the second, such that $|V-W|<\epsilon$ and $|V|,~|W|> R$. Starting from a
solution $Z_1$, $Z_2$ of our equations, we set $Z'_1=Z_1+V$ and $Z'_2=Z_2+W$,
which still obey eq.(\ref{ellpole1}). However eq.(\ref{ellpole2}) is violated at
order $\epsilon$. Choose $Z''_1=Z'_1,~Z''_2=Z''_1-4i\sqrt{g} c_I$ and plug this
in eq.(\ref{ellpole1}). It then gets of order $\epsilon$ but this is an equation
for the variable $Z_1$ which has, by complex analyticity, an exact solution
close to this approximate solution. Taking larger and larger values for $R$ one
gets an infinite number of solutions. Around each of these solutions we have
Puiseux expansions in the variable $\delta t^{1/2}$.

\section{Kowalevski analysis.}

If the swinging Atwood's machine is an algebraically integrable system the
dynamical variables can be expressed algebraically in terms of a linear motion
on some Abelian variety, in particular all variables and time can be
complexified at will. We may expect that, for general initial conditions, the
dynamical variables will blow out for some (in general complex) value $t_0$ of
the time $t$. Around this value the dynamical variables should have Laurent
behavior, hence one expects to find Laurent solutions depending on $N$ parameters
(initial conditions) if the phase space is of dimension $N$. In practice one
searchs for Laurent expansions at $t=0$ (one fixes $t_0=0$) so an admissible
Laurent solution should have $N-1$ parameters, that is 3 parameters for the
example at hand.

The Puiseux solutions we have found in previous section have the following 
singularity:
$x$ and $y$ blow up but $z\to 0$, hence $x^2+y^2 \to 0$. This means that the
singular solutions are such that the mass $m$ goes to the origin but rotating
faster and faster. If we expand $x$ and $y$ in negative powers of $t$ there
must be large cancellations such that $x^2+y^2\to 0$. It is much more
convenient to factorize $x^2+y^2$ and have the cancellation between the two
factors. Reminding that:
$$
x_\pm = x\pm i y
$$
the equations of motion are
\begin{eqnarray}
m \ddot{x}_+ &=& -img+ \lambda x_+ \nonumber\\
m \ddot{x}_- &=&i m g+  \lambda x_- \nonumber\\
M \ddot{z} &=&- M g-  \lambda z \nonumber\\
z^2 &=& x_+x_- 
\label{eqmot}
\end{eqnarray}
The value of $\lambda$ is a consequence of these equations:
$$
\lambda = {mM\over M+m}{ \dot{z}^2 - \dot{x}_+\dot{x}_- +g (y-z) \over  z^2}
$$
where $y=-i(x_+ - x_-)/2$.
Let us remark that this system of equations is invariant under $(x_+,x_-)\to
(-x_-,-x_+)$, in particular $y$ and $\lambda$ are invariant. The system is also
invariant under a similarity transformation:
$$x_\pm(t) \to \mu^2 x_\pm(t/\mu),\quad z(t)\to \mu^2 z(t/\mu), \quad
\lambda(t)\to{1\over\mu^2}\lambda(t/\mu)$$

We first analyze equations~(\ref{eqmot}) at the leading order. We
thus look for solutions of the form:
$$
x_+ = a_1 t^p + \cdots,\quad x_- = b_1 t^q+\cdots,
$$
so that eq.~(\ref{eqmot}) requires
$$
\quad z= c_1 \;t^{p+q\over 2} + \cdots, \quad
c_1^2 = a_1 b_1
$$
At lowest order we then have:
\begin{equation}
\lambda = {mM\over 4(M+m)}{ a_1 b_1(p-q)^2 t^{p+q-2} + 4g (y-z)\over a_1 b_1 t^{p+q}}
\label{eqlambdapq}
\end{equation}
Clearly equations of motion~(\ref{eqmot}) require that $\lambda$ behave as
$1/t^2$ for solutions blowing out as powers. At first sight there are two ways in which this
can happen: when the first term in the numerator is dominant, or when the
second term is dominant. We can always choose $p\leq q$, up to exchange of $x_+$
and $x_-$, hence $p<0$ since we want to have at least one dynamical variable
diverging. On the other hand $z\to 0$ so $q$ is positive, hence $y-z=O(t^p)$.
The first term is dominant when $q<2$, and for $p\neq q$ one has indeed
$\lambda \simeq 1/t^2$. When $q=2$ both terms are of the same
order and for $q>2$ the second term is dominant, so that $\lambda=O(t^{-q})$
which is not allowed. Hence we have basically only two cases to consider, either
$p<0, q<2$ in which the integrable case studied above belongs ($p=-1/2, q=3/2$),
or the case $-2<p<0, q=2$, which, as we will see, covers more general values of
the mass ratio $M/m$.

\subsection{Integrable case.}

Since $p<0, q<2$ we have $p+q-2 < (p,q,{p+q\over 2})$, and we can neglect the
term $g(y-z)$ at leading order in the expression of $\lambda$.
We find, for $p\neq q$:
$$
\lambda=  {mM(p-q)^2\over 4(M+m)}{1\over t^2}+\cdots
$$
Similarly the equations of motion for $x_\pm$ give:
\begin{equation}
p(p-1) =  {M\over 4(M+m)}(p-q)^2 = q(q-1)
\label{pqMm}
\end{equation}
so that $(p-q)(p+q-1)=0$ hence, since $p\neq q$, $p<q$ and we have
$p+q-1 = 0$. Since by positivity in eq.(\ref{pqMm}),
$p$ and $q$ cannot belong to $[0,1]$ this implies, together
with $p>p+q-2=-1$ that:
$$
-1 < p < 0, \quad 1 < q < 2
$$
Using $p+q=1$ the mass ratio takes the form:
$$
M = -4mpq = m[(p-q)^2-1]= m[ (2p-1)^2-1]
$$
and the mass ratio $M/m$ is thus in the interval $]0,8[$.

The integrable case corresponds to $M=3m$, and falls into this
analysis with:
$$
p=-{1\over 2},\quad q = {3\over 2}
$$
These exponents are exactly those we have found in the exact solution of the
elliptic integrable case.
There are no other values of $p$ in $]-1,0[$ compatible with integer values of
the mass ratio $M/m$ which could, according to~\cite{T86}, correspond to
seemingly integrable behaviour. We thus consider, in the following, the
integrable case $M/m=3$.

As noted above the second conserved quantity is given in polar coordinates
for $m=1$, introducing for convenience $H_2=I\,\sqrt{2}/8 $, by:
$$
{1\over 2 \sqrt{2}} H_2 = r^2 \dot{\theta} {d\over dt}{(r \cos (\theta/2))} +
{g\over 2} (r \sin \theta ) (r \cos (\theta/2))
$$
which reads in cartesian coordinates as:
$$
H_2 = {1\over \sqrt{z(z-y)}} (x \dot{y}-y\dot{x} ) {d\over dt} (z^2-z y) + g x 
\sqrt{z(z-y)}
$$
Taking the square to eliminate the square roots, we get:
$$
H_2^2 = {1\over z(z-y)} (x \dot{y}-y\dot{x} )^2\left( {d\over dt} (z^2-z y)
\right)^2 + 2 g x (x \dot{y}-y\dot
{x} ){d\over dt} (z^2-z y)
+ g^2 x^2 (z^2-zy)
$$

We can setup an expansion in powers of $\sqrt{t}$. 
\begin{eqnarray*}
x_+ &=& t^{-{1\over 2}}  (a_1 + a_2 t^{ {1\over 2}}+ \cdots  )\\
x_- &=& t^{3\over 2} ( b_1  + b_2 t^{{1\over 2} } + \cdots ) \\
z &=& t^{{1\over 2} }  (d_1 + d_2 t^{ {1\over 2}}+ \cdots  )\\
\lambda &=& t^{-2}( l_1  + l_2 t^{1\over 2} + \cdots )
\end{eqnarray*}
We already know that
$$
a_1 b_1 = d_1^2,\quad l_1 = {3 m\over 4}
$$

Inserting into the equations of motion, we find the recursive system:
$$ {\cal K}(s)\cdot\pmatrix{a_{s+1}\cr b_{s+1} \cr d_{s+1} \cr l_{s+1} }
=\pmatrix{A_{s+1}\cr B_{s+1} \cr D_{s+1} \cr L_{s+1} } 
$$
$${\cal K}(s) =\pmatrix{m {(s-1)(s-3) \over 4} -l_1 & 0 & 0 & -a_1 \cr
0 & m {(s+1)(s+3) \over 4} -l_1& 0 & -b_1 \cr
0 & 0 & M {(s+1)(s-1) \over 4} + l_1 & d_1 \cr
-b_1 & -a_1 & 2 d_1 & 0 } 
$$
The square matrix  in the left hand side is called the Kowalevski
matrix, and the vector in the right hand side is given by
\begin{eqnarray*}
A_{s+1}&=& \sum_{j=1}^{s-1} l_{j+1}a_{s-j+1} -im g\delta_{s,5} \\
B_{s+1}&=& \sum_{j=1}^{s-1} l_{j+1} b_{s-j+1} +im g \delta_{s,1} \\
D_{s+1}&=& -\sum_{j=1}^{s-1} l_{j+1}d_{s-j+1}-Mg\delta_{s,3} \\
L_{s+1} &=& -\sum_{j=1}^{s-1}  d_{j+1} d_{s-j+1} + \sum_{j=1}^{s-1} a_{j+1}
b_{s-j+1}
\end{eqnarray*}
The determinant of the Kowalevski matrix reads
$$
\det({\cal K}(s)) =- { m^2 d_1^2\over 2}  (s+2)s^2(s-2)
$$
It has a double zero at $s=0$ and a third zero at the integer value $s=2$.
Hence potentially three arbitrary constants may appear in the expansion.
Indeed the miracle happens at the third level where the equations
determining the coefficients $a_3, b_3$ are degenerate, leaving one extra
constant $b_3=c_1$. The rest of the expansion is then completely determined at
all orders. We find in particular:
\begin{eqnarray*}
x_+ &=&  {{d_{1}^2}\over{b_{1}\,\sqrt{t}}}+{{i\,d_{1}^2\,g
}\over{2\,b_{1}^2}}-{{3\,c_{1}\,d_{1}^2\,\sqrt{t}
 }\over{b_{1}^2}}+ {{\left(4\,i\,c_{1}\,d_{1}^2-7\,b_{1}^2\,d_{1}
 \right)\,g\,t}\over{5\,b_{1}^3}}+\\
&&+{{\left(\left(2\,c_{1}\,d_{1}^2+i\,b_{1}^2\,d_{1
 }\right)\,g^2+12\,b_{1}\,c_{1}^2\,d_{1}^2\right)\,t^{{{3}\over{2}}}
 }\over{8\,b_{1}^4}} + \cdots\\
x_- &=& b_{1}\,t^{{{3}\over{2}}}+{{i\,g\,t^2}\over{2}}+c_{1}\,t^{{{5}\over{2
}}}-{{\left(2\,i\,c_{1}\,d_{1}-b
 _{1}^2\right)\,g\,t^3}\over{5\,b_{1}\,d_{1}}}-\\
&& -{{\left(\left(6\,c_{1}\,d_{1}+3\,
 i\,b_{1}^2\right)\,g^2-60\,b_{1}\,c_{1}^2\,d_{1}\right)\,t^{{{7
 }\over{2}}}}\over{40\,b_{1}^2\,d_{1}}}+\cdots
\end{eqnarray*}
 The existence of such a ``miracle'' is exactly what S. Kowalevski noted 
in~\cite{Kowa} for her integrable case of the top. For this to happen one needs
that the determinant of ${\cal K}(s)$ vanishes for the correct number of {\em
integer} values of the recursive variable $s$,
which allows for a new indeterminate to enter the expansion. 
Moreover in this case the linear system has to be solvable which is far from
guaranteed. The {\em general} solution of the equations of motion must admit a
power series expansion, which thus must depend on $2N-1$ arbitrary constants for
a system of $N$ degrees of freedom.  In our
case we find a solution depending correctly on three constants, which extends
the trigonometric solution described above. 

Inserting these expansions into the formula for the energy~(\ref{energy})
we obtain:
$$
E = -{m d_1^2\over 8 b_1^2} ( g^2 + 32 c_1 b_1)
$$
Similarly, the second conserved quantity reads:
$$
H_2^2 = {2 i d_1^5 \over b_1^3} (b_1^2 - 2 i c_1 d_1 )^2
$$

It is interesting to compare these general results to the expansion in the
trigonometric case eqs.(\ref{xplust},\ref{xmoinst}). One finds:
\begin{eqnarray*}
 b_1 &=& e^{-{i\pi\over 4}} {g(K^2+1)\over
4\sqrt{\omega}\sqrt{K}\sqrt{K^2-1}}\\
 c_1 &=& e^{-{3i\pi\over 4}} {g\sqrt{\omega}\sqrt{K^2-1}(K^2+1)\over 32
K^{3\over
	2}}\\
 d_1 &=& e^{-{i\pi\over 4}} {g\sqrt{K}(K^2+1)\over \omega^{3\over
2}(K^2-1)^{3\over 2}}
\end{eqnarray*}
With these values one checks that $H_2=0$ as it should be in the trigonometric
case, and that $H$ is indeed equal to $E$.

\begin{figure}[ht]
\begin{center}
\includegraphics[height= 9cm]{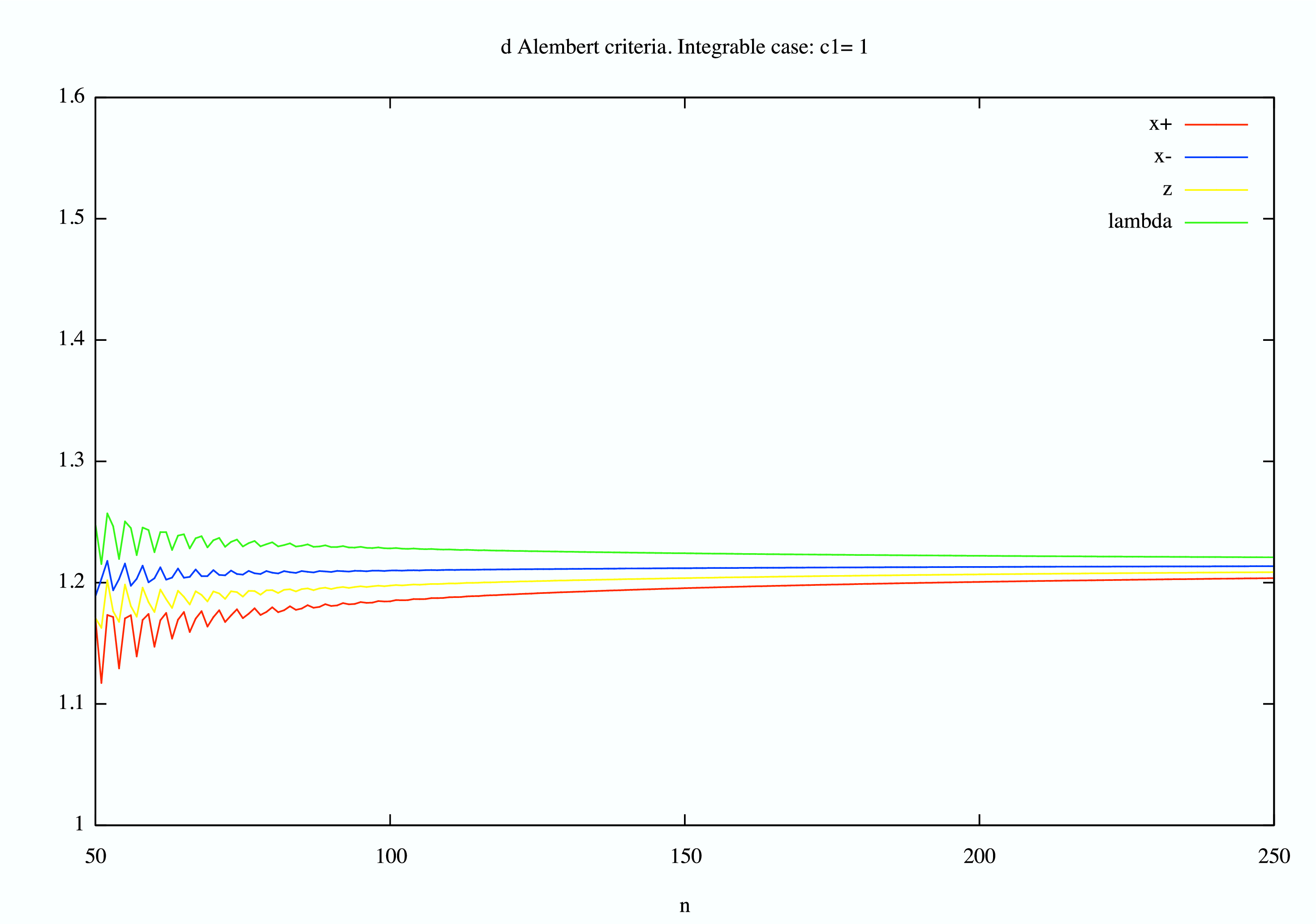}
\caption{d'Alembert criterium for convergence for p=-1/2, q=3/2.}
\label{int_1_DAlembert}
\end{center}
\nonumber
\end{figure}

The dynamical variables $(x,y,z)$ and their time derivatives are expressed in
power series of $\sqrt{t}$. These power series have a non vanishing finite
radius of convergence (we know this at least in the trigonometric case from the
exact solution) and we can check it numerically. To do that we compute the
d'Alembert quotient $|a_{n+1}/a_n|$ relative to a series $\sum_n a_n t^n$
which tends to the inverse of the radius of convergence of this series when it
exists. We present the result of this computation for high order $n$ for the
series $x_+(t)$, $x_-(t)$, $z(t)$, and $\lambda(t)$ in the
figure~(\ref{int_1_DAlembert}).

In this and following similar computations, all values are calculated with
absolute precision rational numbers using a formal computation tool. This
ensures accuracy of the result. 

Since the Kowalevski expansion converges in a
disk, the parameters $(b_1, c_1,d_1)$ appearing in these series, and the origin
of time $t_0$, can be considered as coordinates on an open set of phase space
near infinity~\cite{AVM}. The question then arises to compute the Poisson
brackets in these coordinates.

To do that, we start from:
\begin{equation}
\{A_z(t),x_\pm(t)\} = \pm i x_\pm(t)
\label{azxpm}
\end{equation}
This equation is valid for any time since the time evolution is a canonical
transformation. We thus insert into it the series for $x_\pm(t)$, where
these series are really series in $(t+t_0)^{1\over 2}$. Similarly
$$A_z(t)= i{m\over 2} (x_+\dot x_- - x_- \dot x_+)(t)$$ is expressed as a
series in $(t+t_0)^{1\over 2}$ and eq.(\ref{azxpm}) is an identity in $t$.
The Poisson bracket is computed with the rule:
\begin{eqnarray*}
\{F,G\} &=& \left({\partial F\over\partial t_0}{\partial G\over\partial b_1} -
		{\partial G\over\partial t_0}{\partial F\over
		\partial b_1}\right)\,\{t_0,b_1\}+
	     \left({\partial F\over\partial t_0}{\partial G\over\partial c_1} -
		{\partial G\over\partial t_0}{\partial F\over\partial
				c_1}\right)\,\{t_0,c_1\} +\\
	   &&  \left({\partial F\over\partial t_0}{\partial G\over\partial d_1}
		- {\partial G\over\partial t_0}{\partial F\over
				d_1}\right)\,\{t_0,d_1\} +
	     \left({\partial F\over\partial b_1}{\partial G\over\partial c_1} -
		{\partial G\over\partial b_1}{\partial F\over\partial
				c_1}\right)\,\{b_1,c_1\}+\\
	   &&  \left({\partial F\over\partial b_1}{\partial G\over\partial d_1}
			- {\partial G\over\partial b_1}{\partial F\over\partial
			d_1}\right)\,\{b_1,d_1\}+
	     \left({\partial F\over\partial c_1}{\partial G\over\partial d_1} -
		{\partial G\over\partial c_1}{\partial F\over\partial
					d_1}\right)\,\{c_1,d_1\} 
\end{eqnarray*}
Plugging $F=A_z(t+t_0)$ and $G=x_\pm(t+t_0)$ and identifying term by term in 
$(t+t_0)$ we get an infinite system for the six Poisson brackets of the
coordinates, which is compatible, and whose solution is given by:
\begin{eqnarray*}
\{t_0,d_1\} &=& 0 \\
\{t_0,b_1\} &=& 0 \\
\{t_0,c_1\} &=& {b_1 \over 4 m d_1^2} \\
\{b_1,d_1\} &=& {b_1 \over 2 m d_1} \\
\{c_1,d_1\} &=& {g^2 + 16 b_1 c_1 \over 32m b_1 d_1} \\
\{c_1,b_1\} &=& {g^2 + 32 b_1 c_1 \over 32m d_1^2}
\end{eqnarray*}
We can then check that 
$$\{H,b_1\}=\{H,c_1\}=\{H,d_1\}=0, \quad \{H,t_0\}=1$$
Finally we see that canonical coordinates can be chosen to be the pair 
of couples $(H,t_0)$ and $(\log
b_1,md_1^2)$, hence the Kowalevski constants are essentially Darboux
coordinates in a neighbourhood of infinity.

\begin{figure}[ht]
\begin{center}
\includegraphics[height= 9cm]{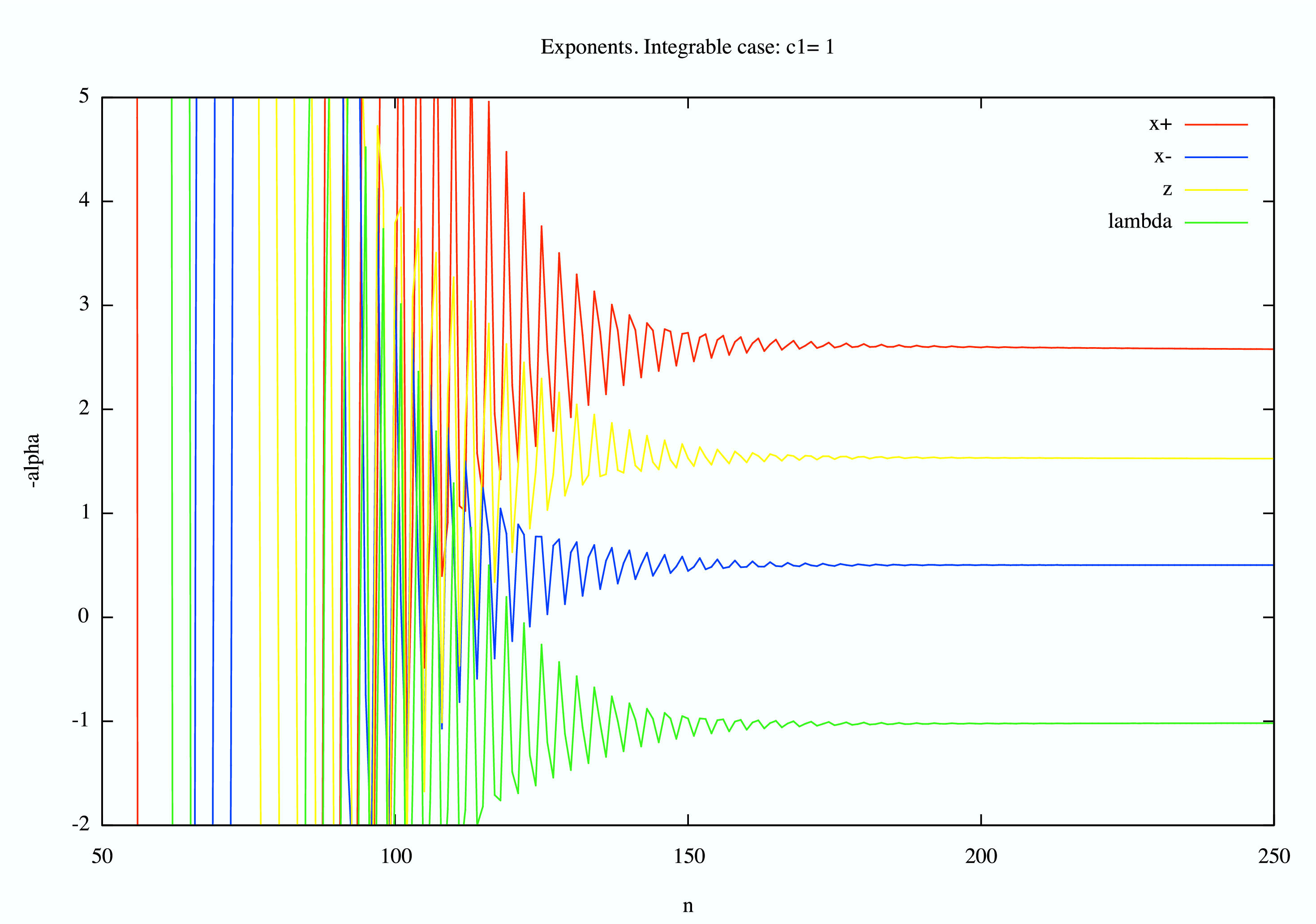}
\caption{Exponents $-\alpha$ as functions of $n$, p=-1/2, q=3/2.}
\label{int_1_Exposants}
\end{center}
\nonumber
\end{figure}

This shows the interest of these Darboux coordinates in a vicinity of
infinity, but the whole question of integrability is a global one. Our problem
is therefore to try to extract some information from the Kowalevski series 
beyond their disk of convergence. In the following we investigate this problem
numerically. First we have seen that $a_{n+1}/a_n$ tends to a complex
number that we call with hindsight $t_\infty^{-1/2}$. Hence $a_n$ behaves
asymptotically as $a_n\simeq t_\infty^{-n/2}$. One can do even better and look
at the prefactor. Assuming that 
$$a_n\simeq A n^\alpha t_\infty^{-n/2}$$
we can extract the coefficient $\alpha$ by computing the quantity:
$$ \lim_{n\to\infty} n^2\left[{a_{n-2}a_n\over a_{n-1}^2}-1\right]=-\alpha$$
We show the result of this calculation in figure~(\ref{int_1_Exposants}).
Note that the curves begin by large oscillations but for $n$ sufficiently
large, in the asymptotic regime, the exponents $\alpha$ tend to constants. 
Comparing with the dominant terms in the binomial formula:
$$\sum n^\alpha z^n \simeq_{z\to 1} (1-z)^{-1-\alpha}$$
we see that setting $z=\sqrt{t/t_\infty}$, we read
from figure~(\ref{int_1_Exposants}) the various exponents:
\begin{center}
\begin{tabular}{rcll}
$x_+(t)$&$\simeq$ &$(1-z)^{3/2},$\\
$x_-(t)$&$\simeq$ & $(1-z)^{-1/2},$&\\
$z(t)$&$\simeq$& $(1-z)^{1/2},$\\
$\lambda(t)$&$\simeq$& $(1-z)^{-2},$
\end{tabular}
\end{center}

The consequence of this observation is that $x_\pm(t)$ have Kowalevski
expansions around $t_\infty$ with indices which are exchanged as compared to 
those around $t=0$. Hence we know that:
\begin{eqnarray*}
x_+ &=&
-b'_{1}\,(t_\infty-t)^{{{3}\over{2}}}-{{i\,g\,(t_\infty-t)^2}\over{2}}-c'_{1}\,
(t_\infty-t)^{ {{5}\over{2}}} +\cdots\\
x_- &=&  -{{{d'_{1}}^2}\over{b'_{1}\,\sqrt{t_\infty-t}}}-{{i\,{d'_{1}}^2\,g
}\over{2\,{b'_{1}}^2}}+{{3\,c'_{1}\,{d'_{1}}^2\,\sqrt{t_\infty-t}
 }\over{{b'_{1}}^2}}+ \cdots
\end{eqnarray*}
where we have introduced a change of sign required by the symmetry
$x_\pm \to -x_\mp$, and the symmetry of the equations of motion under
$t\to t_\infty -t$. The series expansions have new parameters $b'_1$, $c'_1$
and $d'_1$. In the trigonometric case we see from the explicit formulae that
they are equal to the original parameters,  see eqs.(\ref{xplust},
\ref{xmoinst}).

We have learned from the previous analysis that the singularities are always
of the Kowalevski type, with well-defined exponents. This is perfectly
consistent with the exact solution in the trigonometric and elliptic case.

\subsection{Non integrable case.}

We now explore the region of parameters $-2<p<0, q=2$.
We assume that:
$$
x_+\simeq a_1 t^p,\quad x_- \simeq b_1 t^2,\quad z \simeq c_1 t^{{p\over 2}+1}, \quad c_1^2=a_1 b_1
$$
Notice that $z\to 0$ since we assume $p>-2$, and that
$y = -{i\over 2}(x_+-x_- ) \simeq -{i\over 2} a_1 t^p$.
We see that both terms in eq.(\ref{eqlambdapq}) for $\lambda$ contribute:
$$
\lambda \simeq {mM\over M+m} \left( \left({p\over 2}-1\right)^2 -{ig\over 2b_1} \right) {1\over t^2}
$$
The $x_\pm$ equation give:
\begin{eqnarray*}
m p(p-1)&=&{mM\over M+m} \left( \left({p\over 2}-1\right)^2 -{ig\over 2b_1}
\right)\\
2 m b_1 &=& i m g + {mM\over M+m} \left( \left({p\over 2}-1\right)^2 -{ig\over
2b_1} \right) b_1
\end{eqnarray*}
Solving for $b_1$ we find 
$$
M = -4 m {p-1\over p+2}, \quad b_1 = -{ig\over (p-2)(p+1)} 
$$
Notice that the mass ratio is positive if $-2 < p < 0$, and that:
$$
\lambda \simeq {m p(p-1)\over t^2}
$$
For relatively prime integers $r$ and $k$ we set:
$$
p=-{r\over k}, \quad -2k < -r < -k
$$
We perform the Puiseux expansions:
\begin{eqnarray*}
x_+ &=& t^{-{r\over k}}  (a_1 + a_2 t^{ {1\over k}}+ \cdots  )\\
x_- &=& t^2( b_1  + b_2 t^{{1\over k} } + \cdots ) \\
z &=& t^{-{r\over 2k} +1}  (d_1 + d_2 t^{ {1\over k}}+ \cdots  )\\
\lambda &=& t^{-2}( l_1  + l_2 t^{1\over k} + \cdots )
\end{eqnarray*}
We already know that
$$
l_1 = m{r(r+k)\over k^2},\quad a_1 = {d_1^2\over b_1}, \quad b_1 = 
-{ig k^2 \over (r+2k)(r-k)} ,\quad M = 4m {r+k\over 2k-r}
$$
When we plug this into the equations of motion, we get a system of the form:
\begin{equation}
{\cal E}_s:\hskip 1cm
{\cal K}(s)\cdot\pmatrix{a_{s+1}\cr b_{s+1} \cr d_{s+1} \cr
l_{s+1} } = 
 \pmatrix{A_{s+1}\cr B_{s+1} \cr D_{s+1} \cr L_{s+1} } 
\label{equalin}
\end{equation}
where the Kowalevski matrix reads:
$$
{\cal K}(s)=\pmatrix{m {(s-r)(s-r-k) \over k^2} -l_1 & 0 & 0 &
-a_1 \cr
0 & m {(2k+s)(k+s) \over k^2} -l_1& 0 & -b_1 \cr
0 & 0 & M {(k+s -r/2)(s-r/2) \over k^2} + l_1 & d_1 \cr
-b_1 & -a_1 & 2 d_1 & 0 }
$$
and the right hand side of equation ${\cal E}_s$ is given by:
\begin{eqnarray}
A_{s+1}&=& \sum_{j=2}^s a_j l_{s+2-j} -im g\delta_{s,2k+r}\nonumber \\
B_{s+1}&=& \sum_{j=2}^s b_j l_{s+2-j} \nonumber \\
D_{s+1}&=& -\sum_{j=2}^s d_j l_{s+2-j}-Mg\delta_{s,k+r/2} \nonumber \\
L_{s+1} &=& -\sum_{j=2}^s  d_j d_{s+2-j} + \sum_{j=2}^s a_j b_{s+2-j}
\label{nonintrhs}
\end{eqnarray}
For $s=1$, the quantities $A_2$, $B_2$, $D_2$, $L_2$ are meant to be zero.
The determinant of the Kowalevski matrix reads:
$$
\det({\cal K}(s)) =- 6 m^2 d_1^2 { (2k +r) \over k^4(2k-r)} s(s+k)(s-r)(s+k-r)
$$
In order that this determinant vanishes for two positive integer values of $s$,
assuming $k>0$, we should have
$$
r >0, \quad r-k>0
$$
From the equation for $D_{s+1}$ it is natural to choose  $r$ even, otherwise
the weight  $Mg$ would disappear from the problem which is not physical. In
order for $r/k$ to be irreducible, we must choose $k$ odd. Setting $r=2r'$ we
finally get:
\begin{equation}
{k\over 2} < r' < k ,\quad p=-2{r'\over k}
\label{lescaspossibles}
\end{equation}

When things are setup this way the Kowalevski determinant has two strictly
positive integer roots, so that, potentially three arbitrary constants enter the
expansion, or the expansion is impossible. Impossibility occurs when the
right--hand side of equation ${\cal E}_s$ is non vanishing and doesn't
belong to the image of ${\cal K}(s)$ for values of $s$ which are Kowalevski
indices. It turns out that in most cases the right--hand side vanishes as we now
show.

First, since we want to examine the behavior for $s=r-k$ and $s=r$ we can limit
ourselves to studying the system for $s=1,\cdots,r$. In this case the
Kronecker deltas in eqs.(\ref{nonintrhs}) always vanish. For $\delta_{s,2k+r}$
it is obvious, for $\delta_{s,k+r/2}$ note that, since $k\geq 1+r/2$
we have $k+r/2 \geq r+1$.  Since the induction starts 
with $A_2=B_2=D_2=L_2=0$, we get, if $s=1$ is not a Kowalevski index  that 
$a_2=b_2=d_2=l_2=0$, hence the right--hand side vanishes for the next equation
$s=2$. This goes all the way up to $s=r-k$, hence when we hit the
first Kowalevski index, it is always with vanishing right--hand side. The
existence of a non trivial solution $a_{r-k+1},\cdots,l_{r-k+1}$ is thus
guaranteed. Let us assume for the time being that the first Kowalevski
index is such that $(r-k) > 1$, that is $r\geq k+2$.

As a consequence of this previous step, when 
$s=r-k+1$ we find that $A_{s+1}$ reduces to $a_{r+k-1}l_2$ which
also vanishes because $l_2=0$. More generally we have
$A_{s+1}= \sum_{j=r-k+1}^s a_j l_{s+2-j}$ which vanishes when
$s+2-j < r-k+1$ for all $j$ in the sum, and similarly for the other components.
This occurs when $s<2(r-k)$.  For
$s=2(r-k)$  the right--hand side of equation ${\cal E}_s$ doesn't vanish, and
assuming we are not on a Kowalevski index, there is a unique 
non vanishing solution  $a_{2(r-k)+1},\cdots,l_{2(r-k)+1}$. The process
continues and it is easy to show by induction that the right--hand side of
equation ${\cal E}_s$ doesn't vanish only for $s=n(r-k)$, $n$ positive integer,
so non trivial solutions are of the form $a_{n(r-k)+1},\cdots,l_{n(r-k)+1}$. 
Indeed, to get a non vanishing $a_jl_{s+2-j}$ we need to have $j=n(r-k)+1$ and
$s+2-j=n'(r-k)+1$ so that $s=(n+n')(r-k)$. In this case only we have
$A_{s+1},\cdots,L_{s+1}$  and thus $a_{s+1},\cdots,l_{s+1}$,  non vanishing.
This shows that the next non vanishing positions are of the form $(n+n')(r-k)+1$
establishing the recurrence.

The second Kowalevski index is  $s=r$ and this cannot be of the form
$n(r-k)$. Indeed, since $r$ and $k$ are relatively prime, if we have
$r=n(r-k)$ we get $(n-1)r=nk$, hence $n=pr$ and $n-1=qk$ for some integers $p$
and $q$. Then $(n-1)r=nk=qkr=prk$ so that $p=q$ and finally $1=p(r-k)$ 
which is only possible for $p=1$ and $r=k+1$. This is precisely the case we
have excluded up to now. As a consequence, when we arrive at the second
Kowalevski index $s=r$, the right--hand side of equation ${\cal E}_s$
vanishes and there is a non trivial solution, with an extra constant.

We have shown that two new constants of motion 
always appear for all cases $r=k+3,k+5,\cdots,r=2k-2$. This covers an infinite
number of values of the mass ratio $M/m$ for which the Kowalevski criterion is
satisfied (with weak Painlevé solutions), but for which the system is presumably
non integrable.

Finally we discuss the case $k=r+1$. The first Kowalevski index is $s=1$. In
this case the right--hand side vanishes and we have automatically a non trivial
solution $[a_2,b_2,d_2,l_2]$. From this point, all other solutions of the linear
system don't vanish, and in particular, for the second Kowalevski index, $s=r$,
the right--hand side of the system is not trivial. For a solution to exist it
must be in the image of ${\cal K}(r)$. Equivalently, let us consider a covector
$U=[u_1,u_2,u_3,u_4]$ such that $U.{\cal K}(r)=0$. Explicitly:
$$u_1=2\,g^2\,k^5,~u_2=d_{1}^2\,(k+1)(3k+1)^2,~
u_3=i\,d_{1}\,g\,k^2\,(k-1)(3k+1)$$
$$u_4=-2\,i\,mg\,k\,(k+1)(2k+1)(3k+1)$$

The condition to be
satisfied is that the scalar product:
$$W(s)=u_1A_{s+1}+u_2B_{s+1}+u_3D_{s+1}+u_4L_{s+1}$$
of this covector and the right--hand side of eq.~(\ref{equalin})
vanishes for $s=r=k+1$. 

For arbitrary $k$
and $s=3, 4, ...$ we have computed this scalar product $W(s)$, and we have
observed that $W(s)$ has a factor  $(s-k+1)$. For example we get:
$$W(s=3)= - {m c_1^3 d_1^2 (k-2)(k+1)(2k+1)(3k+1)^4
\over 4g^2k^5(k+2)}$$
Note the factor $(k-2)=(r-s)$. For $s=4$ we next get:
$$W(s=4)= {imc_1^4 d_1^2(k-3)(k+1)(2k+1)(3k+1)^4 P_6(k)
\over 96 g^3 (k-1)^2k^8(k+2)^2(k+3)}$$
with the factor $(k-3)=(r-s)$. Here $c_1$ is the Kowalevski constant which
has be introduced at $s=1$, and $P_6(k)$ is some polynomial in $k$ of degree 6.
The factors in the denominator of course come from similar factors in 
$\det({\cal K}(s))$. The expression for $s=5$ has the same type of factors
in the numerator and denominator, with a more complicated polynomial $P_7(k)$
and always a factor $(r-s)$. This behavior is persistent as far as one can
compute. The consequence of the presence of the factor $(r-s)$ is that, for any
$k$, when we arrive at the second Kowalevski index, $s=r=k+1$, the
scalar product $W(k+1)$ vanishes and the linear system is solvable. We can thus
state that for all admissible pairs $(k,r)$ the swinging Atwood machine has weak
Painlevé expansions depending on the full set of parameters.

For example an interesting case occurs when the
mass ratio $M/m=15$ where the system doesn't look chaotic, see~\cite{T185}. This
case is obtained when $k=19$ and $r=26$. The linear system is
solvable in this case, although the new arbitrary constants occur very
far from the beginning of the expansion. We shall refrain to exhibit the
solution in this case, since it is very bulky, and proceed to show what happens
with smaller values of $k$ and $r$.

\subsection{Example: the case $k=3$, $r=4$.}

When $k=3$ we have necessarily $r=4$. The Kowalevski exponents are 
$s=0$, $s=1$, $s=4$. The dynamical variables $x_\pm$ expand in Puiseux series of
$t^{1/3}$ which take the form:
\begin{eqnarray*}
x_+&=& t^{-{4\over 3}} \;d_1^2 \left( {{10\,i}\over{9\,g\,}}+
0 \; t^{1\over 3} +
{{140\,i\,c_{1}^2 }\over{729\,g^3\,}}t^{{2\over 3}}+
{{14000\,c_{1}^3}\over{59049\,g^4\,}}t^{{3\over 3}}+ \right .\\
&& \left . +
{{1960\,i\,c_{1}^4 \,m-32805\,i c_{2}\,g^4}\over{91854\,g^5\,m}}t^{{4\over 3}}
+\cdots \right)\\
x_-&=& t^2 \left(-{{9\,i\,g\,}\over{10}}+
c_{1}\,t^{{{1}\over{3}}}+
{{7\,i\,c_{1}^2 \,}\over{30\,g}} t^{{{2}\over{3}}} +
{{14\,c_{1}^3\,}\over{243\,g^2 }}t^{3\over 3}+ \right . \\
&& \left . +
{{\left(96124\,i\,c_{1}^4\,m-177147\,i\,c_{2}\,g^4\right)}\over{918540\,g^3\,m}} \,t^{{{4
 }\over{3}}}
+\cdots \right)
\end{eqnarray*}
This solution depends on 4 arbitrary constants: $t_0,d_1,c_1,c_2$ (in the above
expansions $t$ should always be understood as $t+t_0$).
We obtain
\begin{equation}
E \equiv H =
{{5\,d_{1}^2\,\left(13412\,c_{1}^4\,m-19683\,c_{2}\,g^4\right)
}\over{91854\,g^4}}
\label{H43}
\end{equation}
The above  constants can be used as local coordinates on phase space. 
To compute the Poisson brackets of the Kowalevski constants, we proceed as in
the previous section considering  $\{ A_z(t), x_\pm(t) \} = \pm i x_\pm(t)$.
\begin{figure}[ht]
\begin{center}
\includegraphics[height= 9cm]{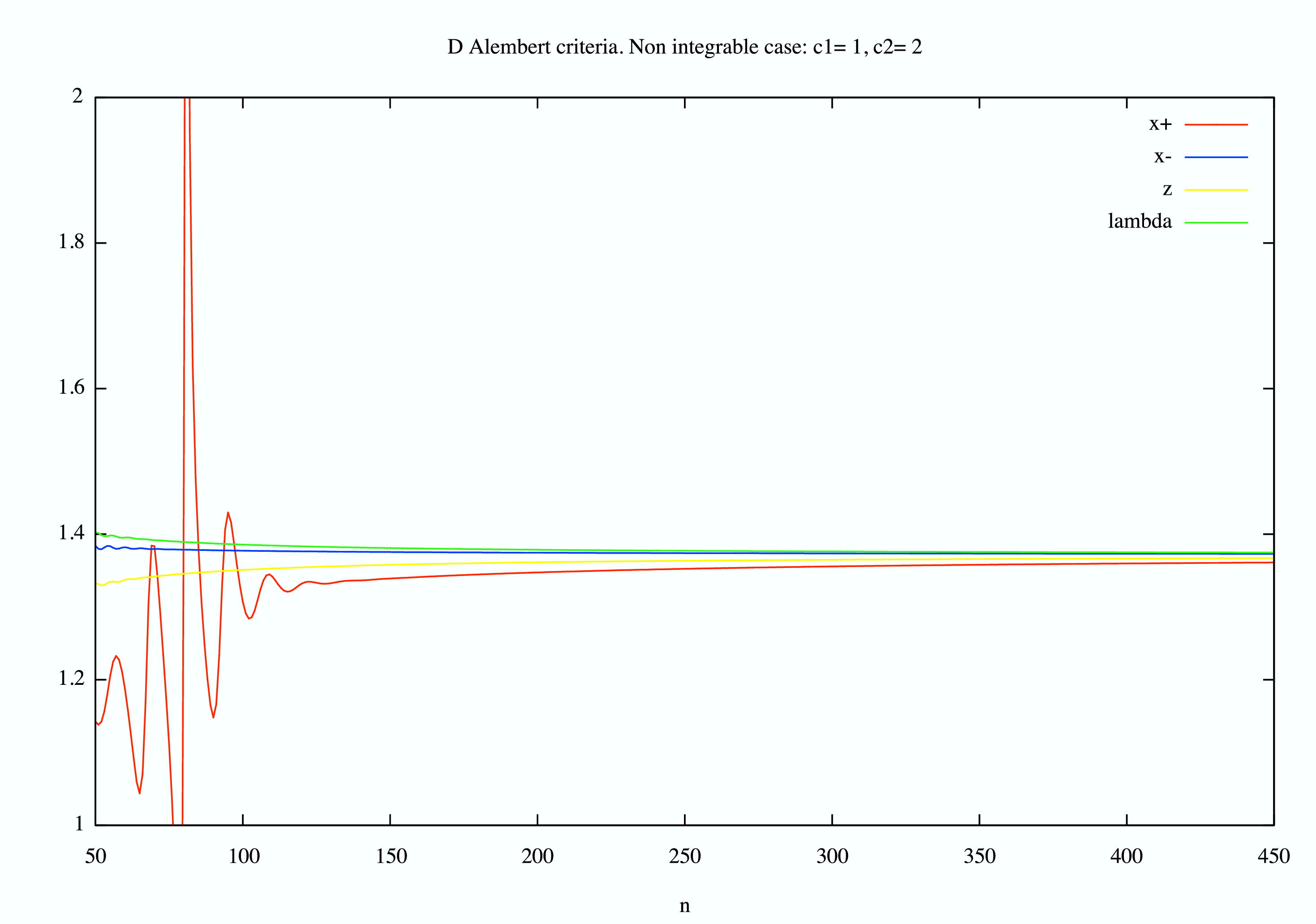}
\caption{d'Alembert criterium for convergence, $\lim a_{n+1}/a_n$
in the non integrable case $c_1=1,~c_2=2$ for $N=450$.}
\label{dalembert_nonint}
\end{center}
\nonumber
\end{figure}
We find:
\begin{eqnarray}
\{ t_0, c_1 \} &=& 0 \label{poi1}\\
 \{ t_0, d_1 \} &=& 0  \label{poi2}\\
 \{ t_0, c_2 \} &=&{{14}\over{15}} \;{1\over d_{1}^2}  \label{poi3}\\
\{ d_1,c_1\}&=&i\; {{3\,g}\over{20\,m}}\; {1\over \,d_{1} }  \label{poi4}\\
 \{d_1,c_2\} &=& i \; {{13412}\over{32805 g^3}}\; {c_1^3\over d_1}  \label{poi5}\\
\{ c_1,c_2 \} &=& -i\; {{\left(13412\,c_{1}^4\,m-19683\,c_{2}\,g^4\right)}\over{65610
 \,d_{1}^2\,g^3\,m}} = -i{7 g \over 25 m} {E \over d_1^4}  \label{poi6}
\end{eqnarray}
It is remarquable that these six relations ensure the compatibility of an
infinite set of relations. One verifies easily
the Jacobi identity in spite of the crazy numbers appearing.
We can compute the Poisson brackets with $H$
\begin{eqnarray*}
\{ H,t_0 \} &=& 1 \\
\{ H, d_1 \} &=& 0 \\
\{ H, c_1 \} &=& 0 \\
\{ H, c_2 \} &=& 0 
\end{eqnarray*}
so that $t_0$ is the conjugate variable of $H$ as it should be and the other
ones are constants of motion. Notice that $(d_1^2, c_1)$ is a pair of canonical
variables commuting with the pair $(H,t_0)$. Kowalevski constants are
essentially Darboux coordinates.

If there were an extra conserved quantity it would therefore be a function
$F(c_1,c_2,d_1)$. The variable $c_2$ can be eliminated through $H$ so that we
can write as well $F(H, c_1, d_1)$. 

\begin{figure}[ht]
\begin{center}
\includegraphics[height= 9cm]{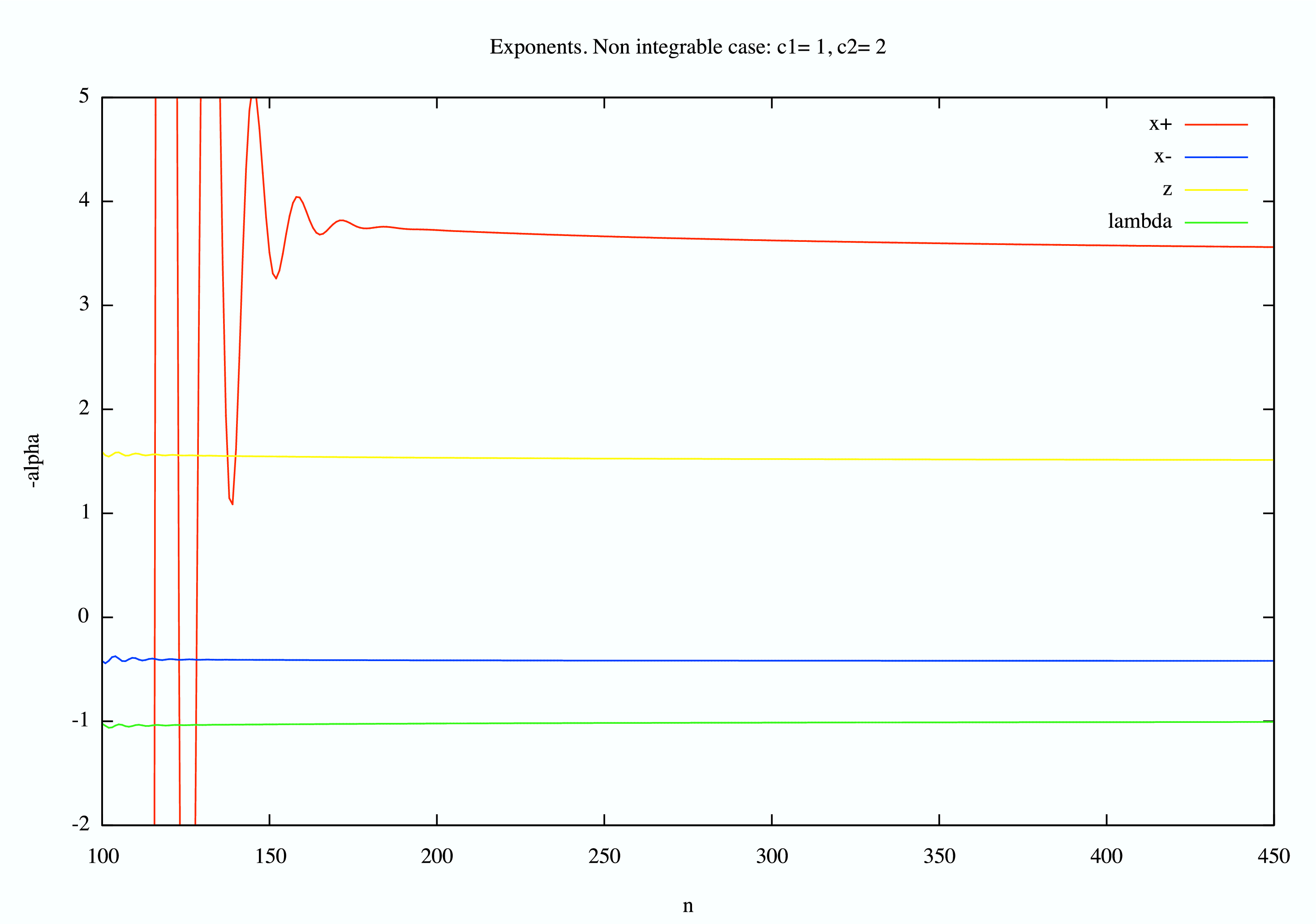}
\caption{Exponents at singularities in the non integrable case $c_1=1,~c_2=2$
for $N=450$.}
\label{exposants_nonint}
\end{center}
\nonumber
\end{figure}

As in the integrable case we can compute numerically the radius of convergence,
and the exponents  which nicely fit with the above Kowalevski analysis, as
shown in Figure(\ref{exposants_nonint}).

\begin{figure}[ht]
\begin{center}
\includegraphics[height= 9cm]{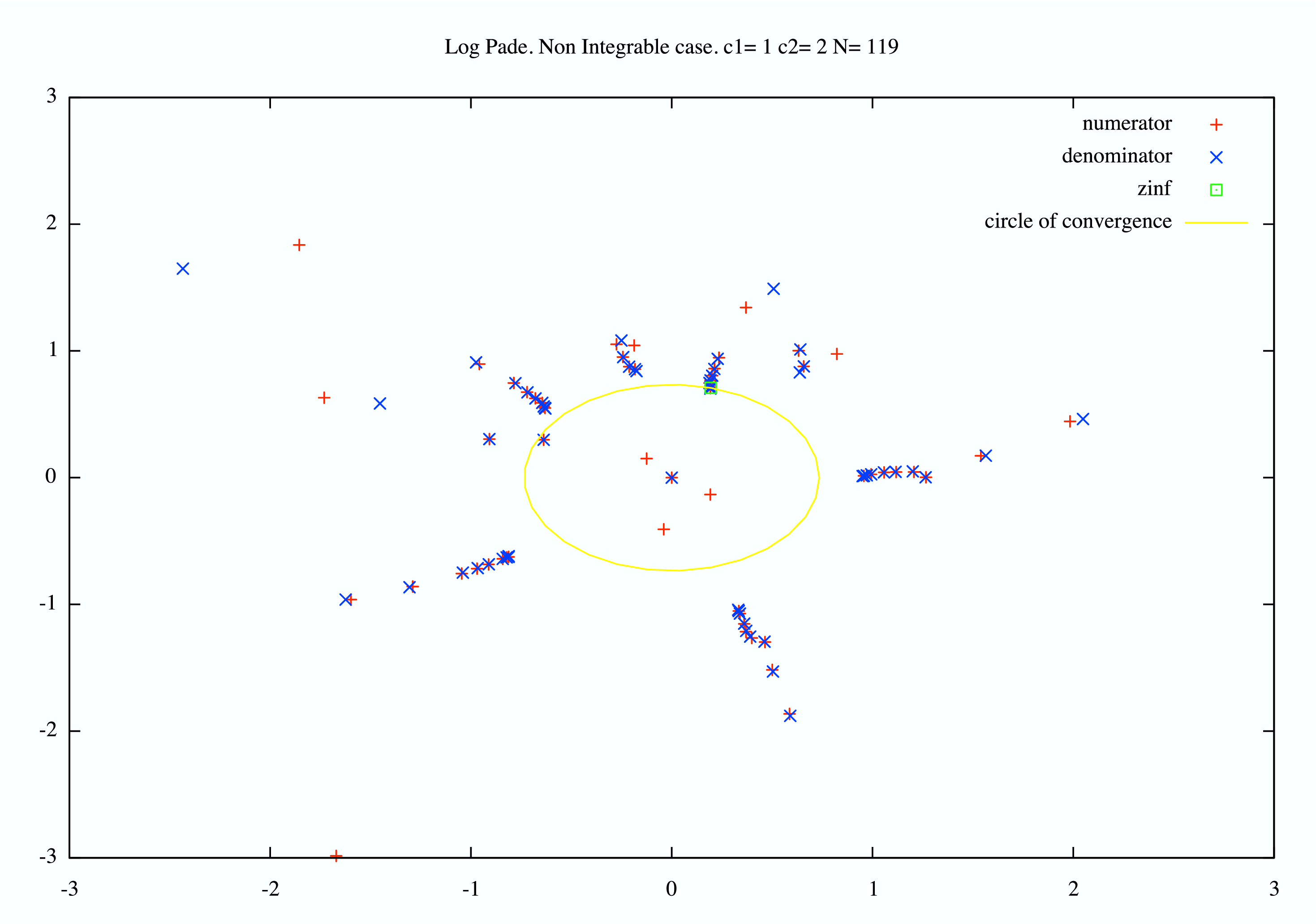}
\caption{Poles and zeroes of Padé approximant $[M,M+1]$ of $\dot{x}_+/x_+$ in
the non integrable case $c_1=1,~c_2=2$, $M=59$, and $N=119$.}
\label{pade_nonint_+}
\end{center}
\nonumber
\end{figure}

To go further, we also compute the Padé approximants of the series. It is
more convenient to consider the logarithmic derivatives $\dot{x}_\pm/x_\pm$
because the residues of the poles are the exponents.
We present the polar decomposition of the $[74,75]$ Padé approximant
fo $\dot{x}_+/x_+$. This shows clearly eight true singularities with residues
respectively -1.33 and 2 (up to numerical errors) consistent with the
Kowalevski analysis. The other poles having small residues
correspond to  strings of poles and zeroes representing algebraic branch
cuts in the Padé analysis. Note we have set $t=z^3$ and we have cancelled the
leading $z^{-4}$ at the origin.
\begin{eqnarray*}
\dot{x}_+/x_+ &=&
 {{2.07+.0366\,i}\over{0.812 +0.618 \, i+z}} +
{{.0295+0.016\,i}\over{0.813 +0.622 \,i+z}}+\cdots\\
&& +{{2.05-.0136\,i}\over{-0.33 +1.04 \,i
 +z}} + {{.0351-.00792\,i}\over{-0.332 +1.04 
\,i+z}}+\cdots \\
&& + {{2.13-.0627\,i}\over{-0.95 -0.012 \,i +z}}+
{{.0725-.0176\,i}\over{-0.954 -1.33 \times 10^{-2}
 \,i+z}}+\cdots\\
&& +{{-1.34-.00154\,i}\over{-0.637 -0.83 \times 
 \,i+z}}
+{{-.00711+0.0025\,i}\over{-6.49 \times 10^{-1}-8.52 \times 10^{-1} \,i+z}}
+\cdots\\
&& + {{2.38-0.027\,i}\over{-0.192 -0.703 \,i+z}} +
{{.0281+.0125\,i}\over{-0.192 -0.705 \,i+z}}+\cdots\\
&& + {{-1.34+3.547 \times 10^{-4}\,i}\over{0.175 -0.84 \,i+z}}+
{{-0.0064-3.344 \times 10^{-4}\,i}\over{0.177 - 0.85 \,i+z}}+\cdots\\
&& +{{2.29+.114\,i}\over{0.629 -0.545 \,i +z}} +
{{0.077-.0335\,i}\over{0.63 - 0.547 \, i+z}}+\cdots \\
&& + {{-1.34-.00401\,i}\over{1.45 - 0.586 \times 10^{-1}
 \,i+z}}+
{{.0802-.0866\,i}\over{1.62 -0.77 \,i +z}} + \cdots
\end{eqnarray*}
We see that this structure is very similar to the one we have observed in the
integrable elliptic case. This semi--local analysis doesn't appear to be able to
discriminate between the integrable and non integrable cases.

\section{Conclusion.}

We have studied the swinging Atwood machine, which is believed to be non
integrable except for the mass ratio $M/m=3$. We have shown on the explicit
solution of the integrable case that the Kowalevski analysis is valid, but
requires weak Painlev\'e expansions. We have extended this weak Painlev\'e
analysis for other values of the mass ratio, and shown that it is valid for
an infinite number of cases. Hence this model is remarkable in that it exhibits
an infinite number of cases where the Kowalevski analysis works at the price of 
using Puiseux expansions. However only one of these cases is known to be
integrable, while the other ones are believed to be not integrable.

In the cases where Kowalevski expansions are available, we have shown that the
constants appearing in these expansions provide Darboux coordinates on an open
set of phase space around infinity. The question of integrability of the system
therefore reduces to the global nature of this coordinate system $(t_0, c_1,c_2,
d_1)$ on phase space.

On this open set, knowing the Poisson brackets eqs.(\ref{poi1}-\ref{poi6}), we 
can try to find the conjugate variable of $t_0$. We find that $H$ must be of
the form:
$$
H = -{15\over 14} d_1^2 c_2 + h(c_1,d_1)
$$
The first term agrees with the exact formula in equation~(\ref{H43}).
The function $h(c_1, d_1)$ is not determined but it is of course crucial to
have a ``good'' function $H(\dot{x}_+, \dot{x}_-, x_+, x_-)$.  Clearly we can,
in principle, invert locally the system of equations
\begin{eqnarray*}
x_+ &=& x_+(t-t_0, c_1,c_2,d_1) \\
x_- &=& x_-(t-t_0, c_1,c_2,d_1) \\
\dot{x}_+ &=& \dot{x}_+(t-t_0, c_1,c_2,d_1) \\
\dot{x}_- &=& \dot{x}_-(t-t_0, c_1,c_2,d_1) 
\end{eqnarray*}
where in the right hand sides we mean the Kowalevski series. In doing so, we will find
\begin{eqnarray*}
t-t_0 &=& T(x_+,x_-,\dot{x}_+,\dot{x}_- ) \\
c_1 &=& C_1(x_+,x_-,\dot{x}_+,\dot{x}_- ) \\
c_2 &=& C_2(x_+,x_-,\dot{x}_+,\dot{x}_- ) \\
d_1 &=& D_1(x_+,x_-,\dot{x}_+,\dot{x}_- ) 
\end{eqnarray*}
but the functions $T, C_1, C_2, D_1$ will behave in general extremely badly. 
All this shows that it is in general impossible to make statements about the
integrability of the system on the only basis of the Kowalevski analysis.  
In this context it is remarkable that the global hamiltonian indeed exists, and
it is even more remarkable that a second global hamiltonian exists in the
integrable case. We see here in a striking way the global nature of
integrability.

In the non integrable case, in an attempt to progress beyond the analysis of a
single singularity, we have used Pad\'e expansions. In this semi--local
analysis, the panorama which appears is still
remarkably similar to the one appearing in the elliptic integrable
case. Hence Kowalevski analysis is not sufficient to characterize integrability.
Nevertheless it is a very non trivial property whose significance remains
mysterious.

\end{document}